\documentclass[review]{elsarticle}
\usepackage{booktabs}
\usepackage{color}
\usepackage{mathrsfs}       
\usepackage{amsfonts}       
\usepackage{mathtools}
\usepackage{array}          
\usepackage{bm}
\usepackage{url}
\usepackage{multirow}

\journal{}

\begin{document}

\begin{frontmatter}

\title{Scalable Double Regularization for 3D Nano-CT Reconstruction}


\author[mymainaddress,mysecondaryaddress]{Wei Tang}

\author[mysecondaryaddress]{Meng Li\corref{mycorrespondingauthor}}

\address[mymainaddress]{Institute of Geology and Geophysics, Chinese Academy of Science, No. 19, Beitucheng Western Road, Chaoyang District, Beijing, China}
\address[mysecondaryaddress]{Department of Statistics, Rice University, 6100 Main Street, MS 138, Houston, TX}

\begin{abstract}
Nano-CT (computerized tomography) has emerged as a non-destructive high-resolution cross-sectional imaging technique to effectively study the sub-$\mu$m pore structure of shale, which is of fundamental importance to the evaluation and development of shale oil and gas. Nano-CT poses unique challenges to the inverse problem of reconstructing the 3D structure due to the lower signal-to-noise ratio (than Micro-CT) at the nano-scale, increased sensitivity to the misaligned geometry caused by the movement of object manipulator, limited sample size, and a larger volume of data at higher resolution. In this paper, we propose a scalable double regularization (SDR) method to utilize the entire dataset for simultaneous 3D structural reconstruction across slices through total variation regularization within slices and $L_1$ regularization between adjacent slices. SDR allows information borrowing both within and between slices, contrasting with the traditional methods that usually build on slice by slice reconstruction. We develop a scalable and memory-efficient algorithm by exploiting the systematic sparsity and consistent geometry induced by such Nano-CT data. We illustrate the proposed method using synthetic data and two Nano-CT imaging datasets of Jiulaodong (JLD) shale and Longmaxi (LMX) shale acquired in the Sichuan Basin. These numerical experiments show that the proposed method substantially outperforms selected alternatives both visually and quantitatively.
\end{abstract}


\end{frontmatter}

\section{Introduction}

Nano-CT is an emerging imaging technique that provides a much higher spatial resolution than its precedent Micro-CT, and has been widely used in many fields such as material science~\citep{Babu2016Resolving}, biomedicine~\citep{Yu2017Phase}, and chemical application~\citep{Cagno2017Combined}. In the area of geoscience, X-ray Nano-CT has attracted growing interests in shale gas and oil as it enables an effective study of the structure of \textit{nanopores}, the dominating pore type in shale that relates to the distribution and capacity of shale gas and oil. High quality image reconstruction in Nano-CT with less artifacts and sharper edges is a crucial step for subsequent analysis (for example, image segmentation~\citep{WANG2019Multiscalecharacterization}), and thus plays an instrumental role in quantitatively analyzing different components such as organic matter, pores, and brittle minerals.
However, the enhanced spatial resolution in Nano-CT poses unique challenges to the inverse problem of reconstructing the 3D structure due to the lower signal-to-noise ratio (than Micro-CT) at the nano-scale, increased sensitive to the misaligned geometry caused by the movement of object manipulator, limited sample size, and a larger volume of data at higher resolution.

Most of the existing methods for CT reconstruction at various scales view the 3D input slice by slice, which might perform well at a high signal-to-noise ratio. While synchrotron radiation makes it possible to reconstruct shale structures at the nano-scale, the noise level in Nano-CT is often inevitably heavier than in Micro-CT or industry-CT, partly due to the geometrical deviation~\citep{sasov2008compensation} and observation system~\citep{kampschulte2016nano} in use. Since the size of Charge Coupled Device (CCD) is much smaller than regular CT devices, the number of projection angles is orders of magnitude smaller than Micro-CT. One motivation of this paper is to utilize the entire data set in CT reconstruction to allow information borrowing within slices as well as between slices, while addressing the daunting memory and computation issues caused by the large volume of data at high spatial resolution.

We propose a scalable double regularization (SDR) method to utilize the entire data set for simultaneous 3D structural reconstruction through total variation regularization within slices and $L_1$ regularization between adjacent slices. This strategy improves upon the traditional methods that build on slice by slice reconstruction at a low signal-to-noise ratio. The proposed SDR method is well suited for a ``blank edge" problem in the projection data of Nano-CT. We develop a scalable and memory-efficient algorithm for routine implementation. The key idea is to exploit the systematic sparsity and some inherent geometry induced by Nano-CT data, without which even loading the data into memory is challenging. Experimental results using simulation and a real data application using shale rocks acquired in the Sichuan Basin show that the proposed SDR often outperforms existing methods both quantitatively and qualitatively. For shale oil and gas, improved reconstruction by SDR is expected to increase the accuracy in analyzing the elementary volume and statistical characteristics of pore sizes, as a result of the sharper edges and less noisy artifacts when recovering vital pore structures. The proposed method is generic and can be applied to other fields using Nano-CT.

\section{Related work}

There has been a rich literature in CT reconstruction since CT was introduced in 1973~\citep{Mostowycz1973Computerized}. Most of the existing methods attempt to inverse a function through its projections recorded at a series of angles. Two main classes of methods include analytic methods such as filtered back projection (FBP) and iterative methods such as algebraic reconstruction technique (ART). Methods based on deep learning have been recently developed to build connections between low-dose CT images and routine-dose CT images through large training data.

\paragraph{Filtered back projection (FBP)}
FBP remains one of the most popular methods in the software of CT devices~\citep{pan2009commercial}. FBP uses analytical solutions based on the Fourier Slice Theorem. Consider a 2D object (such as one slice in a 3D object), projection from each angle gives the value of the object's two-dimensional Fourier transform along a single line. 
The projection data is altered by a high-pass or sharpening filter to preserve sharp edges. Then the final step is backprojection, in which we add together the two-dimensional inverse Fourier transform of each filtered projection~\citep{kak2002principles}. By using the Fast Fourier Transform (FFT) algorithm, the implementation of FBP is very fast. However, FBP does not provide much flexibility to allow the incorporation of prior information on the structure or a model-based approach rooting in the imaging physics~\citep{zeng2014comparison}.

\paragraph{Iterative reconstruction (IR) reconstruction methods.} IR methods are a flexible alternative to improve the reconstruction in the data space iteratively, and have been developed rapidly. Although they always demand more computational power, IR methods usually generate less artifacts than FBP. IR encompasses a variety of methods in the literature from an algebraic perspective, such as algebraic reconstruction technique (ART)~\citep{gordon1970algebraic}, simultaneous ART~\citep{andersen1984simultaneous}, simultaneous iterative reconstruction technique (SIRT)~\citep{Gilbert1972Iterative}, ordered subset simultaneous iterative reconstruction technique (OSSIRT)~\citep{Xu2010On}, multiplicative ART~\citep{Badea2004Experiments}, iterative coordinate descent~\citep{thibault2007three}. Another line of IR methods is from a statistical perspective such as maximum likelihood expectation-maximization~\citep{Lange1984EM} and ordered subset expectation-maximization~\citep{Manglos1995TRANSMISSION}. As an ill-posed problem, CT reconstruction has attracted numerous developments using various regularization to incorporate constraints in the reconstruction~\citep{Chen2013A,sun2014image,kim2016non,chen2014ct,vandeghinste2013iterative,chu2012multi,jia2011gpu,zhang2018regularization,qi2006iterative,thibault2007three,zhang2010bregmanized,wang2012penalized,bai2013low}.

\paragraph{Learning-based methods.} There has been a surge of interest in developing deep learning-based techniques for a variety of tasks in CT reconstruction~\citep{Higaki2019Improvement}. These include using routine CT images to train a neural network to enhance the spatial resolution aiming at the so-called \textit{super-resolution}~\citep{Umehara2017Application,park2018computed}, improving image quality by denoising~\citep{Du2017Stacked,kang2017deep}, reducing the number of projections needed~\citep{Jin2017Deep}, and mapping projection data collected from multi-energy source to monochromatic projection data~\citep{cong2017monochromatic}, just to name a few. Learning-based methods have led to numerous promising results, especially when data are sampled at a limited number of angles as in medical and biological imaging.

Most of the methods mentioned above are motivated by and applied to medical CT, where a typical spatial resolution is around 500$\mu$m~\citep{cnudde2006recent} and the radiation dose on the sample as well as the uncertainty caused by the multi-energy source is always a big concern. In contrast, a synchrotron-based source which is well known for high brilliance, high stability, and high flux, can achieve a super spatial resolution up to nm scale~\citep{kampschulte2016nano}. Moreover, if the sample is rock, it is much less a concern of a high dose in the imaging process than for biological samples. In practice, it is often not realistic to find a large number of training images at the nano-scale to train a neural network for a specific rock task. When the noise level is high, borrowing information from neighboring slices has the potential to improve the accuracy of the reconstructed objects. To this end, we propose a new method that is well suited for Nano-CT reconstruction.

\section{Methods}

\subsection{Background of CT and Notation}
\label{sec_geometry}
We here review the background of CT reconstruction and introduce notation. CT creates an image using X-ray flux measurements from different angles. If the input X-ray photons are mono-energetic, the X-ray intensities passing the object follow the Beer-Lambert Law:
\begin{equation}\label{BLLaw}
  \Delta I/I=-f(x,y)\Delta r,
\end{equation}
where $f(x,y)$ is the linear attenuation coefficient of the object (also called the absorbing function), $\Delta r$ is the small distance that the ray travels, and $\Delta I$ is the attenuation of energy. Estimate the linear attenuation coefficient function $f$ that describes the object's structure is the main task in CT reconstruction. Let $I_{0}$ and $I_1$ be the intensity of the beam entering and exiting the object, respectively, and $\mathscr{L}$ be the path that the X-ray travels. Then Equation~\eqref{BLLaw} yields
\begin{equation} \label{eq:integral}
  I_{1}/I_{0}={\rm exp}\left \{ -\int_{\mathscr{L}}^{}{f(x,y)dr} \right \}.
\end{equation}
Viewed as a mapping from $\mathbb{R}^{2}$ to the set of its line integral, the integral in Equation~\eqref{eq:integral} can be further expressed as the so-called \textit{Radon transform}. Therefore, estimating $f$ boils down to inverting the Radon transform in $\mathbb{R}^{2}$.

\begin{figure}[htb]
  \center{\includegraphics[width=12cm] {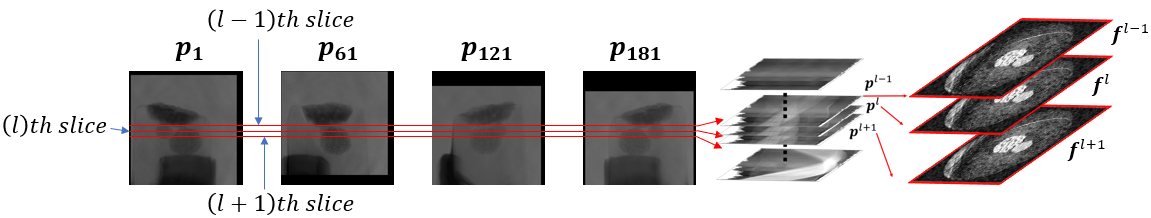}}
  \vspace{-0.2in}
  \caption{\label{pthetaplfl} Demonstration of the image acquiring process in Nano-CT.}
\end{figure}

Synchrotron radiation contains complex devices like collimating mirror and toroidal mirror, which will provide a hollow cone illumination on the sample~\citep{willmott2019introduction}. The scanning geometries can be treated as parallel beam: The parallel light comes from one side of the object, and the detector on the other side get the optical signal by CCD. Iterative reconstruction methods assume the linear attenuation $f(x,y)$ is composed by square grids and in each grid $f(x,y)$ is constant. Suppose we discretize the linear attenuation of the 3D object to an $L \times L \times L$ array, and let $\bm{f}^{l}$ be the $l$th slice where $l = 1, \ldots, L$ from top to bottom of the object. Let $\Theta$ be the set of all projection angles, and $\bm{p}_{\theta}$ be the projection of the object when the object is rotated by $\theta \in \Theta$. Collecting the $l$th row of every $\bm{p}_{\theta}$ leads to the projection of the $l$th slice, denoted by $\bm{p}^{l}$. See Figure~\ref{pthetaplfl} for a demonstration of the introduced notation and the inverse Radon transform mapping each $\bm{p}^{l}$ to $\bm{f}^{l}$. Note that we use the logarithmic scale for all projections $\bm{p}^l$ in this paper.

Throughout the paper, we stack $\bm{f}^l$ and $\bm{p}^l$ column by column into a vector of length $L^2$ and length $L\times|\Theta|$, respectively. Equation~\eqref{eq:integral} leads to the following model
\begin{equation} \label{eq:matrix}
  p_{i}^{l}=\sum_{j=1}^{L^2}{W_{ij} f_{j}^{l}+ \varepsilon_{i}^{l}}, \quad {\text{or in matrix form,}}\quad  \bm{p}^{l}=\bm{W}\bm{f}^{l}+\bm{\varepsilon}^{l},
\end{equation}
where $f_{j}^{l}$ is the $j$th grid in the $l$th slice $\bm{f}^l$, $p_{i}^{l}$ the acquired projection data of the $l$th slice from the $i$th ray, and $\varepsilon_{i}^{l}$ is random noise at the $l$th slice that has mean zero. Here $W_{ij}$ is the contribution of the $j$th pixel to the $i$th ray; for example, if we use a simple project structure, $W_{ij}$  is the intersection length of the $i$th ray and the $j$th pixel. In general, each row of $\bm{W}$ correspond to one ray and each column of $\bm{W}$ corresponds to one pixel.

\subsection{Scalable Double Regularization (SDR)}\label{sec:method}
Exiting methods in Section~\ref{sec_geometry} reconstruct the structure slice by slice, which might perform well at a high signal-to-noise ratio. However, a low signal-to-noise ratio is expected for Nano-CT due to the high resolution in the imaging process and lower photon density~\citep{kampschulte2016nano}. In addition, scalability as well as memory efficiency becomes crucial at high resolution. We propose a scalable double regularization (SDR) approach to reconstruct the 3D Nano-CT structure:
\begin{equation}
\label{eq:objective}
    (\hat{\bm{f}}^{1},\ldots, \hat{\bm{f}}^{L}) = \underset{\bm{f}^1, \ldots, \bm{f}^L}{\rm arg\,min}\sum_{l = 1}^{L}\|\bm{p}^{l}-\bm{Wf}^{l}\|_{2} + \lambda_1 \sum_{l = 1}^L g_{1}(\bm{f}^{l}) + \lambda_2 g_2(\bm{f}^1, \ldots, \bm{f}^L),
\end{equation}
where $g_1$ and $g_2$ are the regularization terms to control the within-slice and between-slice variation, respectively. The estimation in Equation~\eqref{eq:objective} corresponds to a penalized likelihood approach under the assumption that $\bm{\epsilon}^l$ follows a multivariate Gaussian distribution due to the use of the $L_2$ norm. We use the total variation regularization for $g_1(\cdot)$:
\begin{equation}
  g_1(\bm{f}) =: \left\| \bm{f}\right\|_{TV} = \sum_{x, y} \|\nabla f_{x, y}\|_2 =\sum_{x, y}  {\sqrt{(f_{x,y}-f_{x-1,y})^{2}+(f_{x,y}-f_{x,y-1})^{2}}},
\end{equation}
where $\|\nabla f_{x, y}\|_2$ is the Euclidean norm of the image gradient $\nabla f_{x, y}$ at $(x, y)$.

We use the $L_1$ norm of the different images between neighboring slices for $g_2$, i.e., $g_{2}(\bm{f}^1, \ldots, \bm{f}^L) = \sum_{l = 1}^{L-1} \left\| \bm{f}^{l}-\bm{f}^{l-1}\right\|_1$ to favor sparsity between adjacent slices; see Section~\ref{sec:motivation} for a motivation using real data from Nano-CT imaging in tight rock and shale. The use of $g_2$ utilizes the entire dataset for the reconstruction of individual slices, and this information borrowing across slices is particularly well suited for Nano-CT reconstruction where the signal-to-noise ratio is considerably lower than the traditional CT images. We note that other forms involving multiple neighboring slices are also applicable. For example, the summand in $g_2$ may take the form $\|c_0 \bm{f}^l + c_1 \bm{f}^{l - 1} + \ldots + c_t \bm{f}^{l - t}\|_1$ for some $(c_0, \cdots, c_t)$ such that $c_0 + \cdots + c_t = 0$, where the sum to zero constraint encodes the expectation that changes across successive slices are sparse.

We solve the optimization problem in~\eqref{eq:objective} by building upon the gradient descent approach in TV regularization and coordinate descent for lasso~\citep{Tibshirani1996Regression}, which consists of the following steps.

\paragraph{Step 1. lasso using differences of projections} We first optimize~\eqref{eq:objective} focusing on the lasso regularization $g_2$, and address the TV regularization $g_1$ in the next step. Using the linearity in Model~\eqref{eq:matrix}, we take the difference of projections $\bm{p}^{l,l+1} = \bm{p}^{l + 1} - \bm{p}^l$ as input and recast the optimization into a lasso problem:
\begin{equation}
\label{eq:lasso}
{\hat{\bm{f}}}^{l,l+1} = \mathop{\rm arg\,min}\limits_{\bm{f}}\frac{1}{2}\|\bm{Wf}-{\bm{p}}^{l,l+1}\|_{2} + \lambda_2 \|\bm{f}\|_{1},
\end{equation}
where $\hat{\bm{f}}^{l,l+1}$ is an estimate of ${\bm{f}}^{l,l+1} = \bm{f}^{l + 1} - \bm{f}^l$.

We adopt the strategy by~\citep{Friedman2007Pathwise}, and update the $j$th variable of $\bm{f}$ by

\begin{equation}
\hat{f}_{j} \leftarrow S\left(\frac{1}{L\times|\Theta|}\sum_{i=1}^{L\times|\Theta|}W_{ij}\left(p_{i}^{l,l+1}-\sum_{k\neq j}^{L^2}W_{ik}\hat{f}_{k}\right),\lambda_2\right),
\end{equation}
where $S(\zeta, \eta)$ is the soft-thresholding operator given by
\begin{equation}\label{GGG_3}
S(\zeta,\eta) = \left\{
\begin{aligned}
\zeta - \eta &\quad {\rm if} \ \zeta >0 \ {\rm and} \ \eta<|\zeta|,\\
\zeta + \eta &\quad {\rm if} \ \zeta <0 \ {\rm and} \ \eta<|\zeta|,\\
\qquad0 &\quad \text{otherwise.}
\end{aligned}
\right.
\end{equation}
At given $\lambda_2$, we use zero as the initial values and estimate $\hat{\bm{f}}^{l,l+1}$ iteratively until convergence.

\paragraph{Step 2. TV regularization} Based on the estimates of the differences of slices, we then use a gradient decent approach in TV regularization to reconstruct the 3D object. To this end, we approximate  the gradient of $\left\| \bm{f}\right\|_{TV}$ by
\begin{equation}\begin{split}
    \frac{\partial \left\| \bm{f}\right\|_{TV}}{\partial f_{x, y}} & \approx
    \frac{(f_{x,y}-f_{x-1,y})+(f_{x,y}-f_{x,y-1})}{\sqrt{\epsilon+\|\nabla f_{x, y}\|_2}}\\
    & -\frac{f_{x+1,y}-f_{x,y}}{\sqrt{\epsilon+ \|\nabla f_{x + 1, y}\|_2}}
    -\frac{f_{x+1,y}-f_{x,y}}{\sqrt{\epsilon+ \|\nabla f_{x, y + 1}\|_2}},
\end{split}\end{equation}
where $\epsilon$ is a small positive number to ensure numerical stability~\citep{Laroque2008Accurate}. We use Barzilai-Borwein (BB) step~\citep{barzilai1988two} to calculate the step size. The initial values are specified by the Kaczmarz method, i.e.,
\begin{equation}\label{eq:Kaczmarz}
  \bm{f} \leftarrow \bm{f}-\frac{\alpha \left\{ \left \langle \bm{f},\bm{W}_{i} \right \rangle -\bm{p}_{i} \right\} \bm{W}_{i}}{\left\|\bm{W}_{i}\right\|_2^{2}},
\end{equation}
where $\alpha$ is the relaxation factor and $\bm{W}_{i}$ is the $i$th row of $\bm{W}$. At each iteration, the reconstruction is updated by combining the construction of slices and differences of slices as follows
\begin{equation}
\begin{aligned}
\label{eq:average}
\hat{\bm{f}}^{l} \leftarrow (\hat{\bm{f}}^{l}+\hat{\bm{f}}^{l-1}+\hat{\bm{f}}^{l+1}+\hat{\bm{f}}^{l,l+1}-\hat{\bm{f}}^{l-1,l})/3.
\end{aligned}
\end{equation}

The algorithm terminates when a certain convergence criterion is met. In particular, we terminate the algorithm when the relative change of the reconstruction falls below a threshold, or the number of iteration reaches a specified maximum.

In all our experiments, we set the relaxation factor $\alpha$ to be 1. At a chosen slice, $\lambda_1$ is tuned by an L-curve method~\citep{Hansen1993Theuse} using the first two terms in Equation~\eqref{eq:objective}. We find that the proposed method is not sensitive to the choice of $\lambda_1$ as long as it falls into a reasonable range. The selection of $\lambda_2$ can be carried out through cross-validation as usually implemented in the lasso. To speed up the algorithm, we recommend to alternatively search $\lambda_2$ at a limited number of grid points and choose a value that leads to reasonable reconstructions on selected slices.

\subsection{Scalability and Memory efficiency}
For real data with a $512 \times 512$ resolution and 180 angles of projections, the dimension of $\bm{W}$ is around $ (512 \times 512) \times (512 \times 180) =  (2.6 \times 10^{5}) \times (9.2 \times 10^{4}) > 2 \times 10^{10}$ pixels, leading to daunting memory and computation issues. A dense matrix with the size of $\bm{W}$ may occupy 180Gb memory, exceeding the maximum memory limit by many workstations. However, in the CT imaging process, one ray can only go through few pixels of the entire image: for a $m \times n$ matrix $\bm{W}$ only $\sqrt{n}$ of each row of $\bm{W}$ is non-zero. We exploit this extreme sparsity to store $\bm{W}$'s efficiently by only storing the non-zero values and their locations, which dramatically reduces the memory demand. For example, we only need less than 1.75Gb memory using the example above.

The sparse structure of $\bm{W}$ not only provides efficient memory allocation for the output, but also improves the computation by taking advantage of efficient sparse matrix operations. The process of solving Equation~\eqref{eq:Kaczmarz} and lasso requires calculating many inner products between $\bm{W}$ and other vectors.  In the solvers for the TV and lasso regularization, the utilization of sparsity reduces the complexity of each inner product operation from $O(n)$ to $O(\sqrt{n})$.

The Matlab code to implement the proposed method with demonstration is available at \url{https://github.com/xylimeng/SDR-CT}.

\section{Simulation}
\label{sec:synthetic data}
\begin{figure}[htb]
  \center{\includegraphics[width=\linewidth] {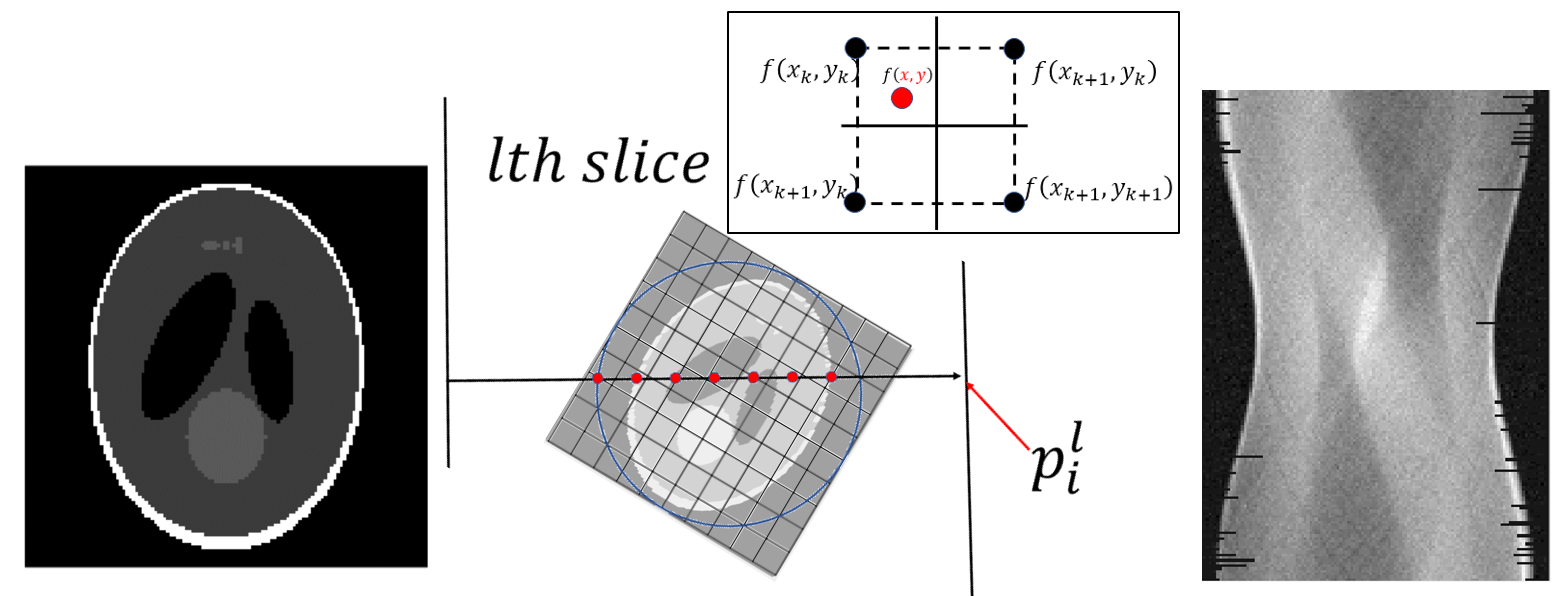}}
  \caption{\label{numerical_setup} Phantom of 64th slice, Demonstration of generating $\bm{f}^l$ ($l  = 64$) via bilinear interpolation, and Simulated projection corrupted by Gaussian noise and blank edges(from left to right).}
\end{figure}

\begin{figure}[htb]
  \center{\includegraphics[width=0.9\linewidth] {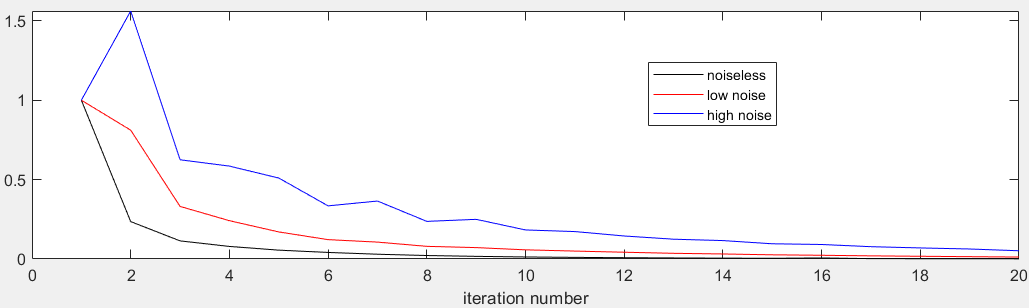}}
  \caption{\label{Convergence}Convergence of SDR under various scenarios: noiseless (black), low noise (red), and high noise (blue). The $x$-axis is the iteration number; the $y$-axis is the relative change at each iteration.}.
  \vspace{-0.12in}
\end{figure}

In this section, we demonstrate the proposed method using a classic 3D shepp-logan phantom of size $128\times 128\times 128$. For each slice, we sample equal-spaced points on each ray line. The value of $f(x, y)$ at any $(x, y) \in \mathbb{R}$ is determined by bilinear interpolation according to its closest sampled points:
\begin{equation}
  f(x,y)\approx \begin{bmatrix}
   x_{k+1}-x & x-x_{k}
  \end{bmatrix}
  \begin{bmatrix}
   f(x_{k},y_{k})   & f(x_{k},y_{k+1}) \\
   f(x_{k+1},y_{k}) & f(x_{k+1},y_{k+1})
  \end{bmatrix}
  \begin{bmatrix}
   y_{k+1}-y\\
   y-y_{k}
 \end{bmatrix}\label{bblinear}
\end{equation}
In the middle of Figure~\ref{numerical_setup} demonstrates such discretizaton process to generate $\bm{f}^l$. We consider three scenarios depending on the standard deviation $\sigma_{\text{true}}$ of the Gaussian noise added to the true projection: noiseless ($\sigma_{\text{true}} = 0$), low noise ($\sigma_{\text{true}} = 0.5$), and high noise ($\sigma_{\text{true}} = 1$). In order to mimic the blank edges as in the real data (see Section~\ref{sec:motivation} for more details), we randomly remove some pixels of the projection on the edge. The number of pixels removed is the same for the three noise levels. The right plot in Figure~\ref{numerical_setup} shows projection of 64th slice corrupted by Gaussian noise($\sigma_{\text{true}} = 1$) and blank edges.
We compare the proposed SDR with FBP, OSSIRT (one of the best IR methods without regularization~\citep{Xu2010On}), and TVART~\citep{Chen2013A} (one of the best IR methods combined with regularization). 
We use the method described in Section~\ref{sec:method} to tune $\lambda_1$ and $\lambda_2$ for SDR. In particular, we choose the middle slice to select $\lambda_1$ from 0 to 3, leading to $\lambda_1 = 0.5$; we search $\lambda_2$ from 0 to 0.04, and the selected values are 0.005 (noiseless), 0.015 (lowe noise), and 0.03 (high noise). We use the same $\lambda_1$ for the method of TVART. We run up to 20 iterations for all methods to have a fair comparison, when they appear to stabilize. Figure~\ref{Convergence} shows the relative $L_2$ change of the reconstructed structures, indicating that the proposed SDR converges fast. It uses 20 iterations to achieve a relative change of 0.1$\%$ (noiseless), 1.2$\%$ (low noise), and 5.2$\%$ (high noise).

In order to assess the performance of each method, we use the signal-to-noise ratio (SNR) and the structural similarity (SSIM) index~\citep{Wang2003Multiscale} between the reconstructed structure $\hat{f}$ and the true structure $f$:
\begin{equation}
   \text{SNR} =10 \log \left( \frac{\sum_{i=1}^{L^2}{\left( f_{i}-\mu_{f}\right)^{2}}}{\sum_{i=1}^{L^2}{\left(\hat{f}_{i}-f_{i}\right)^{2}}} \right),
\end{equation}
\begin{equation}
\text{SSIM} =\frac{\left( 2\mu_{f}
   \mu_{\hat{f}}+C_1\right) \left( 2\sigma_{f\hat{f}}+C_2\right)}{\left( \mu_{f}^{2} +\mu_{\hat{f}}^{2}+C_1\right) \left( \sigma_{f}^{2}+\sigma_{\hat{f}}^2 +C_2\right)},
\end{equation}
where $\mu_{\hat{f}},\mu_{f}$ are the means of $\hat{f}$ and $f$, $\sigma_{f},\sigma_{\hat{f}}$ are the standard deviations of $\hat{f}$ and $f$, and $\sigma_{f\hat{f}}$ is the covariance of $\hat{f}$ and $f$.
In addition, we use the contrast-to-noise ratio (CNR)~\citep{Timischl2014TheCNR,Welvaert2013Onthe} to evaluate the reconstruction relative to region of interest (ROI), which is defined as
\begin{equation}
    \text{CNR}=\frac{|\mu_{target}-\mu_{background}|}{\sqrt{\sigma_{target}+\sigma_{background}}},
\end{equation}
where the subscript ``target" and ``background" indexes pre-specified ROI and background, respectively. For example,
in the Shepp-Logan phantom experiment the background is the gray areas inside the white boundary but outside ellipses, and the target region is the small gray ellipses above the two black ellipses in Figure~\ref{numerical_setup}.

\begin{figure}[htb]
  \center{\includegraphics[width=0.95\linewidth] {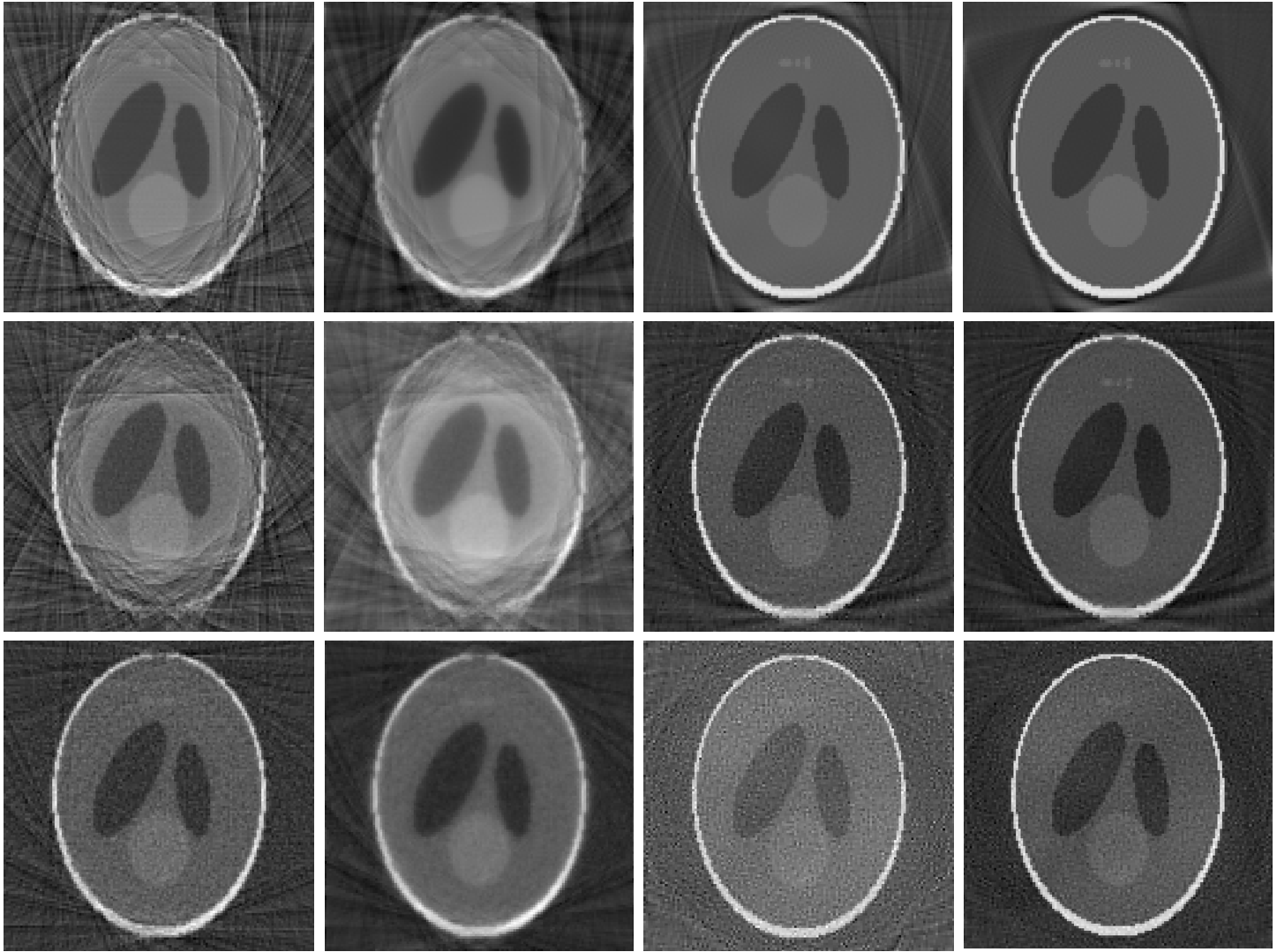}}
  \caption{\label{Reconstructed} Reconstruction of the 64th slice using FBP, OSSIRT, TVART, and SDR (left to right). The three rows correspond to the cases of noiseless, low noise, and high noise.}.
  \vspace{-0.12in}
\end{figure}

Figure~\ref{Reconstructed} shows that there are significant artifacts in the reconstructed images of FBP and OSSIRT in the noiseless case (see the first row), suggesting that they suffer considerably from the blank edge issue. TVART and SDR provide the best visualization consistently across various scenarios, but the proposed SDR tends to give clearer reconstruction at a high noise level (3rd row). Since all iterative methods use the same number of iterations, the comparison between OSSIRT (blurred) and SDR (clearer) indicates the advantage of incorporating regularization in the reconstruction.

\begin{table}
	\setlength{\tabcolsep}{4pt}
	\centering
	\label{table1}
	\caption{Comparison of different methods using 3D Shepp-Logan phantom.}
	\begin{tabular}{cccccccccc}
		\toprule
		\multirow{2}{*}{Methods} & \multicolumn{3}{c}{noiseless} & \multicolumn{3}{c}{low noise}& \multicolumn{3}{c}{high noise}\\
			\cmidrule(lr){2-4} \cmidrule(lr){5-7}
		\cmidrule(lr){8-10}
		& SNR & SSIM & CNR & SNR & SSIM & CNR & SNR & SSIM & CNR\\
		\cmidrule(lr){1-1} \cmidrule(lr){2-4} \cmidrule(lr){5-7}
		\cmidrule(lr){8-10}
		FBP & 9.21 & 0.74 & 0.82 & 6.69 & 0.66 & 0.69 & 15.23 & 0.88 & 0.71\\
		OSSIRT & 8.90 & 0.71 & 1.00 & 6.30 & 0.62 & 0.84 & 14.61 & 0.85 & 0.66\\
		TVART & 24.41 & 0.95 & 2.51 & 21.64 & 0.94 & 1.22 & 12.12 & 0.87 & 0.70\\
		CNR & $\bm{28.2}$ & $\bm{0.97}$ & $\bm{3.19}$ & $\bm{25.69}$ & $\bm{0.96}$ & $\bm{1.53}$ & $\bm{23.4}$ & $\bm{0.95}$ & $\bm{0.98}$\\
		\bottomrule
	\end{tabular}
\end{table}

Table~\ref{table1} quantitatively assesses each method by reporting SNR and SSIM averaged across the 20 slices in the middle and CNR at the 64th slice. We use these slices as they contain most of the important structures for the 3D Shepp-Logan phantom data. We can see that SDR is uniformly the best method across all metrics and in all scenarios. The superior performances of the proposed method suggest that the implemented double regularization utilizing the entire dataset substantially improves the reconstruction.


\section{Real data application}
\label{sec:real data}
In this section, we apply the proposed method to two main types of Nano-CT experiments: absorb-contrast experiments and phase-contrast experiments. In particular, we use real Nano-imaging datasets of an absorb-contrast Jiulaodong (JLD) shale sample and a phase-contrast Longmaxi (LMX) shale sample. In absorb-contrast experiments X-ray imaging captures intensity changes where we only need to invert the absorption of the sample to the light, while in phase-contrast experiments phase information is additionally recorded in the projections. Propagation-based phase-contrast imaging (PPCI) extends the possibilities of X-ray absorption imaging especially when the contrast of attenuation coefficients between different regions is low~\citep{Cloetens1997Observation}.

\subsection{Empirical evidence: sparsity and blank edge}
\label{sec:motivation}
We first use the absorb-contrast JLD shale sample to provide empirical evidence that aligns with the motivation of the proposed method SDR. Our absorb-contrast JLD shale experiment uses samples sourced from well W201, a key well in the Sichuan Basin that produces industrial-scale gas flow~\citep{zongqing2012shale}.
The Nano-CT experiments were conducted at beamline BL01B~\citep{song2007x}, the National Synchrotron Radiation Research Center (NSRRC) in Hsinchu, Taiwan, which provided 2D micrograph and 3D tomography with a pixel size of 50nm. This dataset has been previously analyzed  by~\cite{wang2016characterization}, where they have detected four types of pore structures in JLD shale (i.e., organic matter pores, interparticle pores, intraplatelet pores within clay aggregates, and intercrystalline pores within pyrite) and found Nano-pores are the primary contributing pore structure to understanding this shale reservoir. Note that our real data application has a radically different focus, which is to improve the 3D structure reconstruction particularly for shale rocks at the nano-scale. The samples are milled to cylinder to produce 180 sequential image frames at equal-spaced angles.

\begin{figure}[htb]

  \center{\includegraphics[width=12cm] {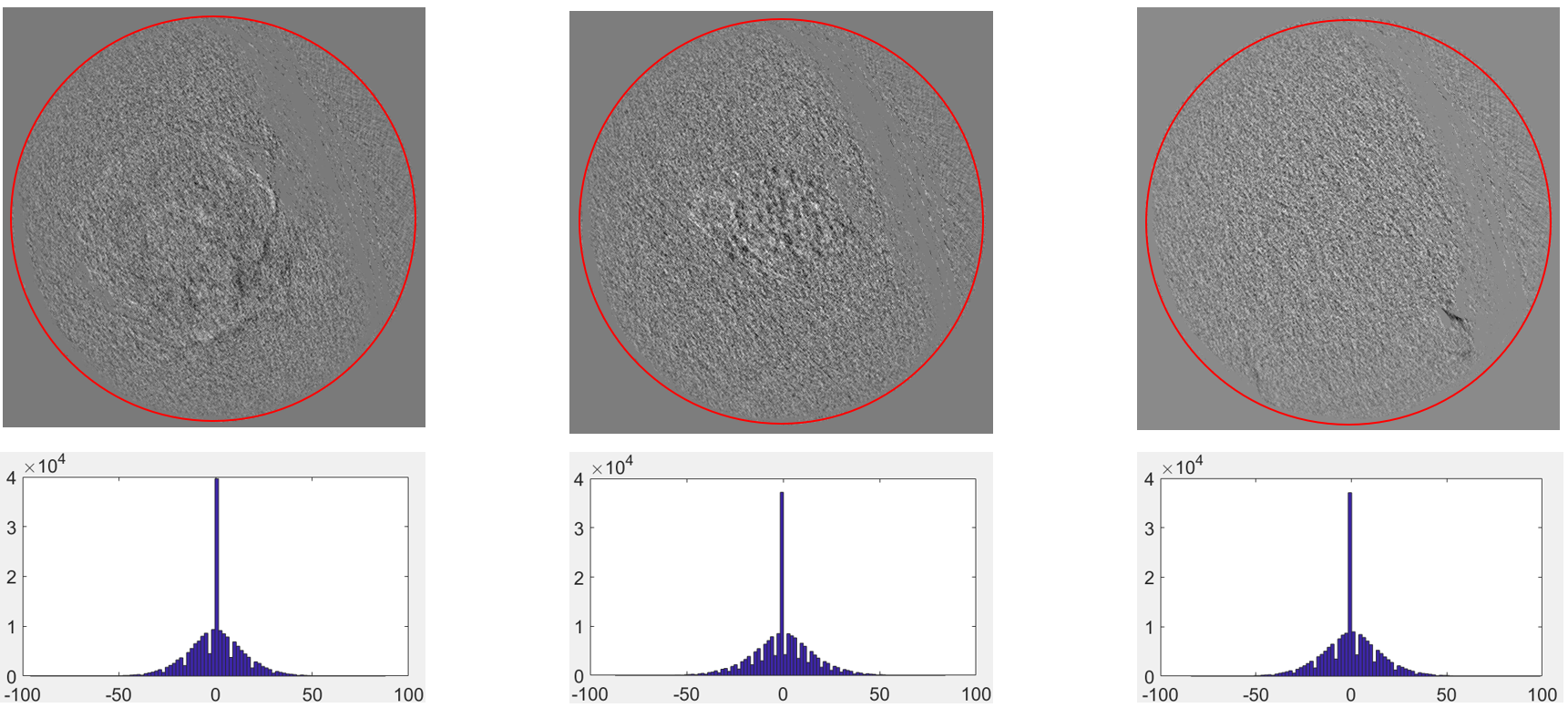}}
  \caption{\label{diff_real} The difference between two adjacent slices given by FBP by subtracting the $l$ slice from the $l + 1$th slice. The three columns correspond to $l = 200, 300, 400$, left to right. The second row is the corresponding histogram of intensities within the red circle.}
\end{figure}

We apply the FBP algorithm to all image frames. The field of view in Nano-CT should be composed of a few pores and high-density material. Their absorption coefficient to light varies greatly. It is expected that the difference between adjacent slices should be mostly zero but contain edges of the structure, i.e., sparse. Figure~\ref{diff_real} shows the difference of selected adjacent slices after reconstruction along with the histograms of the intensity, which clearly confirms the sparsity pattern. A slice by slice algorithm such as FBP does not enforce sparse differences, and to incorporate such constraints into the reconstruction procedure is expected to improve accuracy and sharpness of edges.

\begin{figure}[htb]
  \center{\includegraphics[width=12cm] {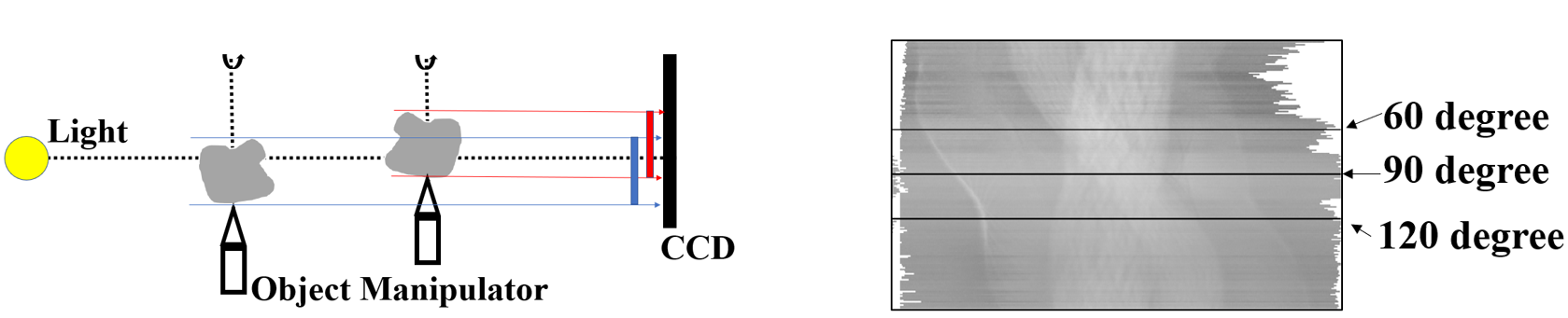}}
  \caption{\label{fig:blank edges} geometric deviation and projection of 280th slice corrupted by blank edges }
\end{figure}

Another characteristic of Nano-CT imaging data is that there are blank edges in the raw projection data. Blank edges are more often in Nano-CT because even a tiny movement of the object becomes significant compared to CT at coarser scales (such as nm). 
The cartoon in the left plot of Figure~\ref{fig:blank edges} demonstrates that if the object manipulator moves up a little, then after alignment the bottom of the projection data cannot be recorded. The same phenomenon happens when the object manipulator moves left or right. The vibration changes irregularly with the angle, causing most of the projection $\bm{p}_{\theta}$ to lose some data on edge after the geometric correction. The right plot of Figure~\ref{fig:blank edges} shows such an effect using one slice from our observed real data.

FBP and traditional IR methods essentially assume uniform sampling without considering blank edges in the reconstruction. Moreover, the adjacent projections $\bm{p}^l$ in middle slices (which contain the most useful information about the object) have identical blank edges, according to the device setup in image acquiring. This systematic geometry strongly supports information borrowing across slices, as a simultaneous reconstruction across slices should alleviate the missing data issue of blank edges.

\subsection{Comparison of reconstructions}
We compare FBP (which is applied by most Nano-CT devices directly) with our algorithm SDR using the aforementioned laboratory measurements of JLD shale and LMX shale. We use the same method as in Section~\ref{sec:synthetic data} for parameter tuning when implementing SDR, and select $\lambda_1 = 1$ and $\lambda_2 = 0.03$ in both applications.

\subsubsection{Absorb-contrast JLD shale sample}
\begin{figure}[!htb]
  \center{\includegraphics[width=0.95\linewidth] {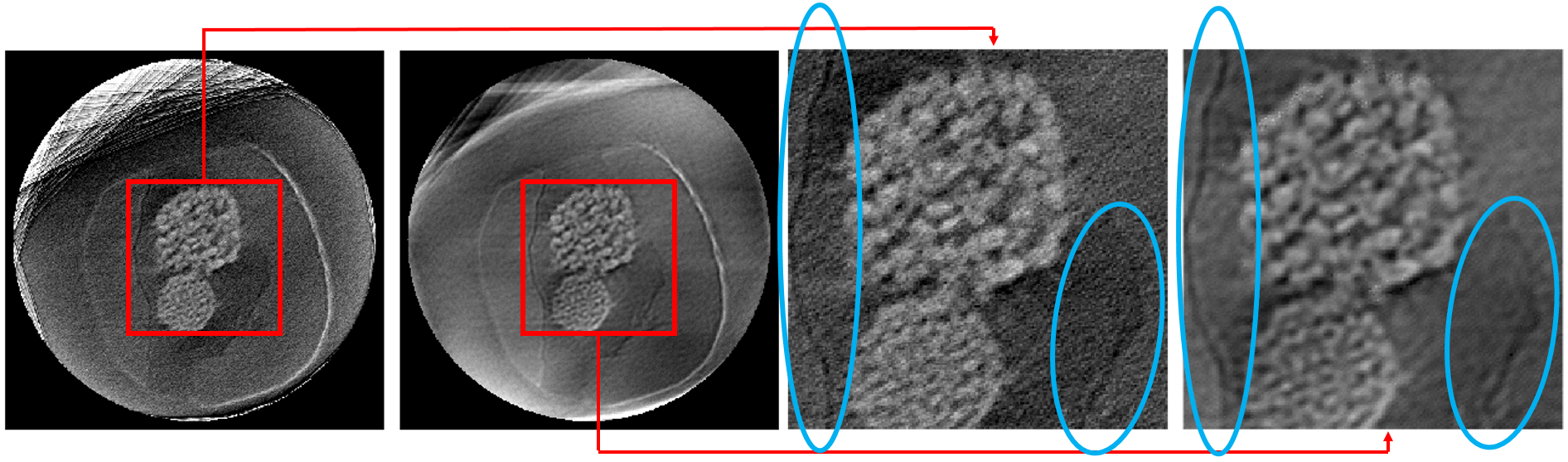}}
  \caption{\label{real_data} Comparison of SDR and FBP using a JLD sample. From left to right is the reconstruction of FBP, SDR, and zoomed plots of the region within the red frame by FBP and SDR. Blue circles in the zoomed plots highlight a long microfracture and a short microfracture.}
\end{figure}
The visual comparison is shown in Figure~\ref{real_data}.
We find that the proposed SDR exhibits less noisy artifacts than the FBP reconstruction. In the zoomed plots, framboid pyrite and intercrystalline pores are shown between a ``longer microfracture" of dozens of micrometers (the long black line enclosed by the blue circle on the left) and a  ``shorter microfracture" (enclosed by the blue circle at the right bottom of zoomed plots). The result given by SDR provides a much clearer and shorter microfracture than FBP, especially in the middle of the shorter microfracture where the reconstructed fracture of FBP is hardly distinguishable from noise.

\begin{figure}[!htb]
  \center{\includegraphics[width=0.95\linewidth] {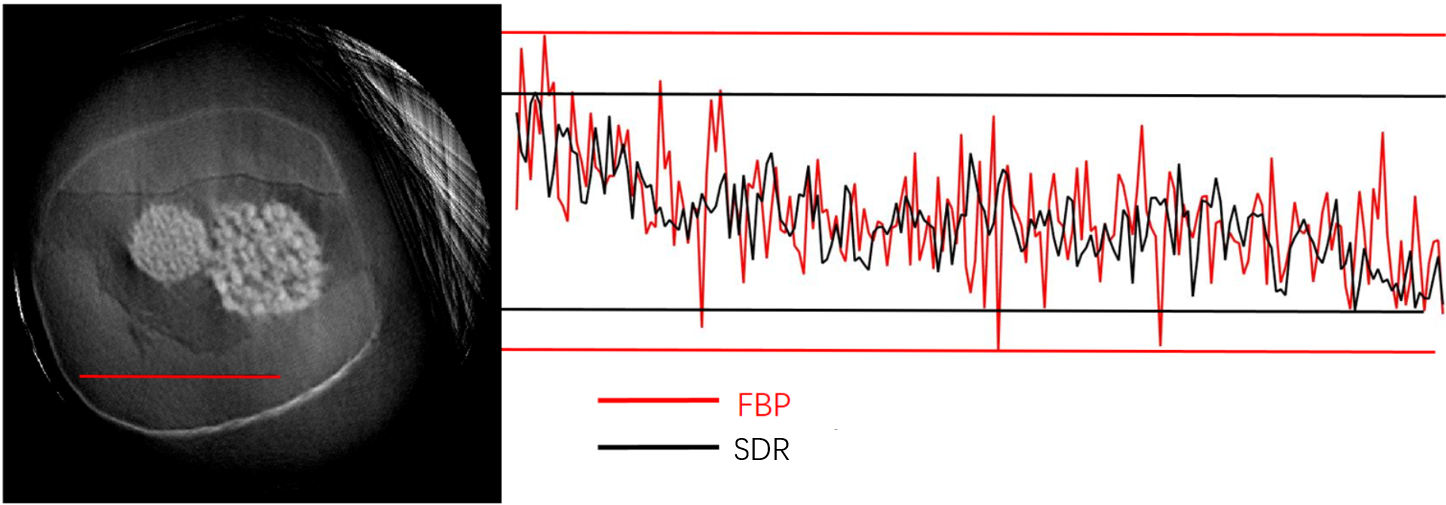}}
  \caption{\label{abs_profile} Profile of one line in the reconstructions of FBP and SDR.}
\end{figure}

In Figure~\ref{abs_profile}, we compare FBP and SDR focusing on the profile of one line marked in the left plot that is expected to be close to constant. We can see that the reconstructed profile by FBP has much more fluctuation and a wider range than the one by SDR, indicating that SDR gives smoother reconstruction when there are no edges present.

\begin{figure}[!htb]
  \center{\includegraphics[width=0.9\linewidth] {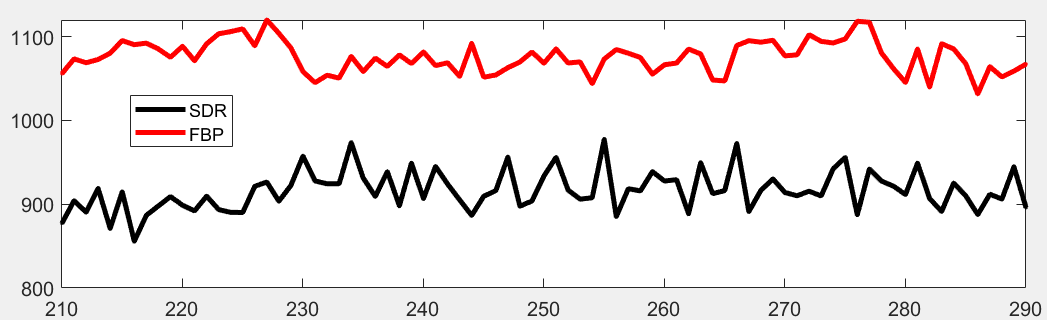}}
  \caption{\label{PLOT2} NRSS of SDR vs. FBP from the 210th slice to the 290th slice of JLD shale.}
\end{figure}

Since the ground truth is not available in real data application, the metrics used in Section~\ref{sec:synthetic data} can not be calculated. We instead calculate a no-reference structural sharpness (NRSS) metric to measure the sharpness of the reconstruction quantitatively:
$$
    \text{NRSS}  =\sum_{x}^{}{\sum_{y}^{}{\{ \left( f_{x+1,y}-f_{x,y}\right)^{2}+\left( f_{x,y}-f_{x,y+1}\right)^{2}\}}}.
$$
Figure~\ref{PLOT2} compares the NRSS from the 210th slice to the 290th slice, clearly indicating that SDR provides much sharper reconstructions.

\subsubsection{Phase-contrast LMX shale sample}
The sample in this section is LMX shale in the Sichuan basin. LMX shale in the Sichuan basin is by far the best-studied shale gas reservoir in China~\citep{wang2019multiscale}. Previous studies have shown that the main pore size ranges from 2nm to 500nm~\citep{Zhangqin2015,wang2019multiscale}. We extract phase information using the software Phase-sensitive X-ray Image processing and Tomography Reconstruction (PITRE)~\citep{Chen2012PITRE}, then apply FBP and the proposed SDR to achieve CT reconstruction.

\begin{figure}[!htb]
  \center{\includegraphics[width=0.65\linewidth] {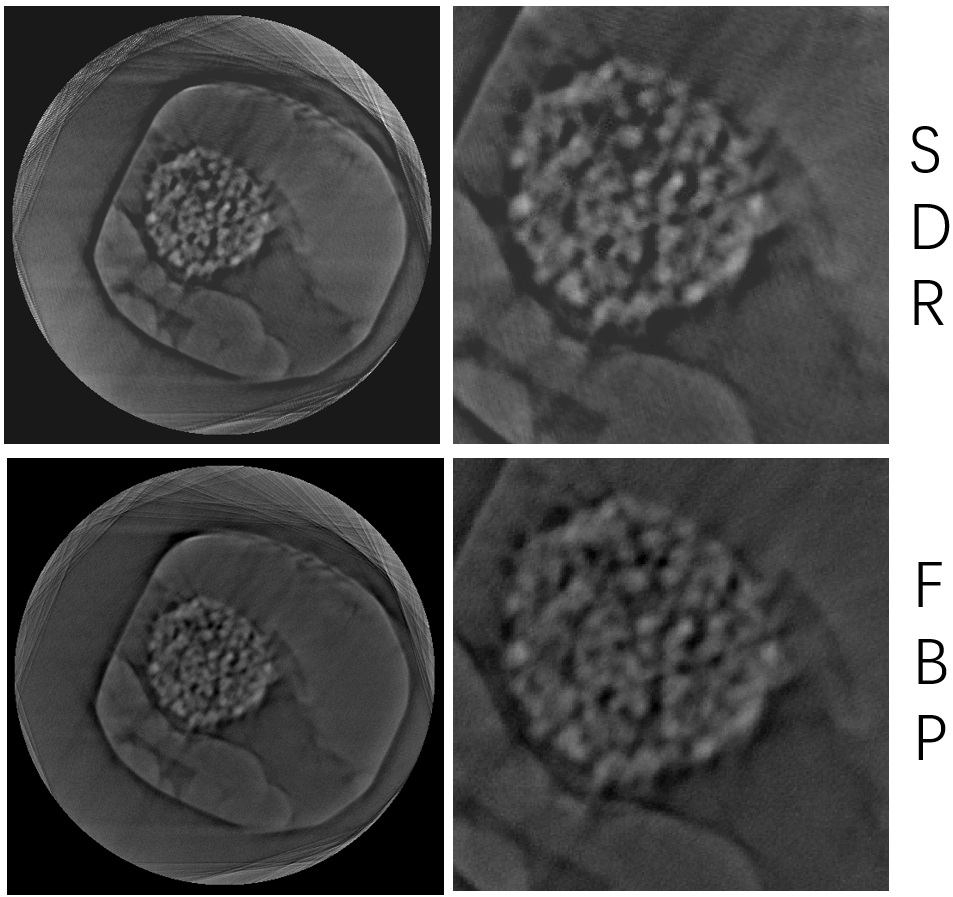}}
  \caption{\label{Real_shizhu} Comparison of SDR and FBP using an LMX sample.}
\end{figure}

\begin{figure}[!htb]
  \center{\includegraphics[width=0.9\linewidth] {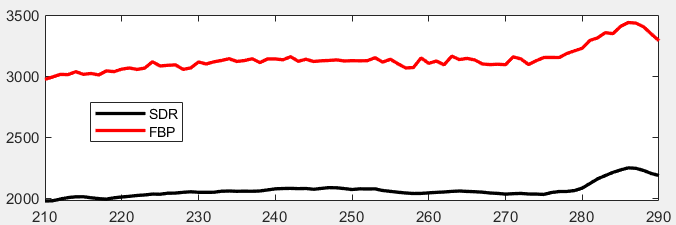}}
  \caption{\label{NRSS2} NRSS of SDR vs. FBP from the 210th slice to the 290th slice of LMX shale.}
\end{figure}

Figure~\ref{Real_shizhu} plots the reconstructed results and zoomed images by SDR (upper plots) and FBP (lower plots). From the visual comparison, the reconstruction given by SDR provides sharper edges with enhanced contrast. In addition, Figure~\ref{NRSS2} compares the NRSS from the 210th slice to the 290th slice, which confirms that SDR provides much sharper reconstructions.

The enhanced contrast by SDR is expected to improve the subsequent post-processing steps such as segmentation. In practice, such improved reconstruction by SDR is expected to increase accuracy in analyzing the elementary volume and statistical characteristics of pore sizes, pyrite, and organic matter, partly due to the sharper edges and less noisy reconstructions when recovering vital structures.

\section{Conclusion}
\label{sec:conclusion}

Nano-CT is a non-destructive method capable of providing nm scale information in a shale gas reservoir. A fast reconstruction method that allows information borrowing across slices is desirable to tackle unique challenges in Nano-CT reconstruction, including a low signal-to-noise ratio and a large volume of data. To this end, we propose a scalable algorithm for 3D structural reconstruction for simultaneous Nano-CT reconstruction utilizing the entire dataset with regularization both within slices and between slices, as opposed to slice by slice reconstruction in most of the existing methods. We achieve memory-efficiency by exploiting the systematic sparsity and geometry in Nano-CT imaging. The developed algorithm is motivated by and applied to real datasets in nano-CT imaging of shale samples acquired in the Sichuan Basin. Our numerical experiments show that the proposed method provides sharper edges and less noisy artifacts, both visually and quantitatively. The enhanced construction by the proposed method is expected to improve the subsequent post-processing steps and increase accuracy in analyzing the elementary volume and statistical characteristics of pore sizes. The Matlab code is made publicly available at Github for routine implementation by practitioners. 

\section*{Acknowledgements}
The authors thank Prof. Yanfei Wang for providing the shale sample data used in the real data application. Wei Tang is partially supported by a scholarship from the China Scholarship Council and Strategic Priority Research Program of the Chinese Academy of Sciences (Grant No. XDB1002010). Meng Li is partially supported by an ORAU Ralph E. Powe Junior Faculty Enhancement Award and grant 1R24MH117529 of the BRAIN Initiative of the United States National Institutes of Health. The work was conducted when the first author visited the Department of Statistic at Rice University.
\section*{References}

\bibliography{mybibfile}

\end{document}